# The Effect Of Online-Cooperative Homework On Students' Academic Success

Semseddin Gunduz
Computer Education and Instructional Technologies
Necmettin Erbakan University Faculty of Education
Konya, Turkey
semsedding@gmail.com

Aysen Namlu
Istanbul Commerce University
Istanbul, Turkey
agurcan@ticaret.edu.tr

*Abstract*— In this study, the effect of online –cooperative learning homework practices on academic success of students is searched. The experience group of the research consists of 58 students from Anadolu University Education Faculty Education of Computer and Instruction Technology Section. Students in A section are taken to traditional method by neutral appointment; those in B section are taken to online homework practice method. In each class consisting of 29 people, it's decided that 14 students prepare their homework individually; the rest 15 students prepare their homework with cooperative as triple groups. It's students' own choice to prepare their homework individually or cooperatively. There has been a success scale at the end of the teaching period. According to research results, there isn't statistically considerable difference between students who attend traditional homework practices and online homework practices. According to research results, there isn't statistically considerable difference between students who attend individual homework practices and cooperative homework practices. The academic success of the students who attend online-based individual homework practices is higher than traditional individual and online based cooperative learning homework practices.

I. INTRODUCTION (*Heading 1*)

Increasing of information of the subjects to be learnt and not having time to complete the new titles, enhancement of requests to self-renew and improve themselves increased the importance of educational activities outside the classroom. The activities planned should be considered as a support to education process or an alternative, not as out of class education. One of the important out of class activities is homework. Today, it seems impossible for the teachers to provide the participation of all students in overcrowded classes in education process. Teachers can achieve the participation of students assigning them homework out-of-class.

Many studies are done on whether the homework is effective in education process or not. Reference [1] advocates that ones who defend the assigning homework state homework has an important role in academic life and improve the academic success. In contrast, Reference [2] indicates that assigning a lot of homework affect the success negatively.

Reference [3] says that he had the most portion of conflict with his students on homework and in order to make them complete their homework and like the homework, he tried many systems. As a solution, he asked his students to do their homework out-of-home. As a result of technological developments, at present, students have the chance to complete their studies out-of-home. They also have the chance to receive, complete and send their homework from different places (laboratories, mobile phones and internet cafes operated by various institutions) with internet connection. The web pages for monitoring asynchronous courses, there should be a page for homework and exercises [4]. Universities using internet- based distance learning online courses benefit homework in teaching-learning process and when evaluating the students they consider the homework [5].

Reference [6] due to lack of face to face interaction with the students using the web-based training systems are working alone and probably they have less support by classmates and feel under pressure. To overcome this problem, teachers should practice activities based on cooperative assignments in web-based teaching systems. There are five basic elements of cooperative learning: positive interdependence, face-to-face interaction, individual scalability, social skills and group process [7]. These can be applied in cooperative based practices as in the following format:

First, students in cooperative based homework group will achieve together or fail together. Secondly, students in this group should always be in contact with each other. Third, the aim of cooperation is not only to bring out homework all together, but also to contribute to the homework and to acquire future ability to accomplish the homework alone. Fourth, students' social skills should be improved in order to continue the process effectively. Fifth, students should come together sometimes to discuss how the process is going on and find solutions for the problems.

Reference [8] in their study "Internet Homework Activities and Traditional Homework Activities: the Effects on Achievement, Completion Time, and Perception" develops a post-test group design. As a result, it is indicated that both homework methods increase the motivation of students. However, no significant difference in exam success of these two-homework methods is found.

Reference [9] in their study conducted with 171 junior high school students with disabilities take part in either in the group based on cooperative homework or individual one. The result of evaluation is found in favour of the group based on cooperative homework in terms of homework completion rate and accuracy degree of homework. No significant difference was found between students' successes in mathematics.

Homework issue is being studied for a long time but the effects of teacher during the homework process is focused on less [10]. Researchers evaluating the effects of homework on success haven't consistent results. One of the reasons is thought to be due to the fact students participate in different homework practices. Research data collected by the online homework and collaborative planning and implementation are expected to be helpful. Discovering the future teachers' skills in preparing homework seems important to take necessary precautions. Research is expected to contribute and light the future studies.

The purpose of this research can be expressed as" is there any significant difference between the academic success of students participating in different practices?"

According to this aim, study is going to try to find solutions to the following questions:

• In academic success of students participating online homework practices and traditional homework practices ,

• In academic success of students participating individual homework practices and cooperative homework practices ,

• In academic success of students participating four different practices (traditional individual, online individual, traditional cooperative, online cooperative)

Is/are there a significant difference?

## II. METHOD

In this study, in order to determine how different homework practices affect the academic success, 2 x 2 factorial design is used. The first factor modifies homework variable and includes two different styles, traditional and online methods. The second factor modifies the homework technique variable and includes two different styles, individual and co-operative. In the study, the effects of homework method and technique on the student success, dependent variable of the study.

*A. Participants*

The study is conducted with 58 students attending Programming Language I Classes at Anadolu University, Faculty of Education, Department of Computer Education and Instructional Technology BTO201 in the first term of 2004-2005. As an objective way, students in A class take traditional method and in B class take online homework method. In each class, there are 29 students, 14 of them are asked to do their homework individually, and 15 of them are asked to do their homework based on cooperatively. Each group has set a name for themselves. Online cooperative group also creates an e-group on Yahoo using their group name. In the group, the teachers and the other group members are assigned as moderators. %40 of homework grade is given for group activities while the assessment process. For the group study grades, imagery records and e-group usage are considered.

*B. Data and Collection*

In the study, midterm exam, final exam and homework grades are considered to evaluate the academic achievement of students. The success grade of students in the study is distributed as follows: Midterm Exam: 30%; Final Exam: 40%; Homework grading: 30% (in three parts, 9%, 9% and 12%). Exam questions are prepared by researchers and previous instructors of the course. Exam papers are evaluated by these instructors and the results are compared with the results of the same experts. The correlation analysis result is found as .9955.

Before starting the research, participants are informed about the preparation of homework, homework practices and then samples are shown and questions of students are replied in 2 class hours. Subject is explained the students using traditional methods during their class hour. Students are assigned to do homework including all the subjects two weeks before their exam and one week later, their papers are collected back. Experimental group follows the rules below:

Traditional individual homework practice: Homework is assigned to the students during the class time orally and the published materials are shown as a source. Students, in order to ask questions and have some assistance have the opportunity to meet the professor face-to-face for two class hours (90 min) in office and they are encouraged. Students have the chance to consult their professor whenever he is free. They give their homework during the class hour. Prepared papers are presented in the class the next hour. The mean of the scored given by the students affect the homework grade by %40 and remaining %60 is completed by the professor. Students are given feedbacks and their grades in the next class hour.

Traditional cooperative homework practice: The same process as in traditional individual homework is followed. However, students in the group searched the subject given as homework individually. After completing their searches, group members gather and record their cooperation, document sharing, discussion and task distribution activities as voice or video and give this document to the professor. Group members arrange a document about tasks, task distribution and contribution by the members. Homework is presented to the professor as a report. %40 of homework grade is given to the students for their activities. Image records are examined one by one.

Online individual homework practice: Homework has been declared in course instructor's website and sent to the e-mail addresses of students. When homework be delivered, how to be delivered, what should be paid attention while preparing homework, homework success score effect, teaching the lecture schedule, sources that can used while doing the homework and online resources for reading are announced on the same website again and also with electronic mail. In order to have assistance whenever they need and for online help professor's e-mail is given and students are encouraged. Students have the chance to chat with each other and the professor for two-class hour during the week. Microsoft Messenger is used as an online communication tool. Through this program, students ask their questions to the professor and receive an instant feedback. By using this program, while preparing their homework students use "white board" and "share" activities through this program. At the end, students send their homework to the professor through e-mail and present it to the other students. The mean of the scored given by the students affect the homework grade by %40 and remaining %60 is completed by the professor. Students learn their homework results and feedbacks within 24 hours through e-mail.

Collaborative online homework practice: Same process is followed as in online individual homework practices. However, the group of students, the task before them as an individual investigated. However, students in the group searched the subject given as homework individually. After completing their searches, group members discuss their communication, collective work; document sharing, discussion and task distribution activities as in e-group. Group members share their documents with their group friends and explain the unclear points online. Communication of group members with each other, collective work, document sharing, discussions and task distributions of students are saved electronically. %40 of homework grade is given to the students for their group activities. For this purpose, e-group records are examined individually. Homework is presented as one report online.

*C. Data Analysis and Interpretation*

In the study, the arithmetic mean, standard deviation, correlation analysis, analysis of variance and Fisher's LSD test are used. As significance level, .05 is found on confidence level. All the statistical analysis are done using SPSS 10.0 (Statistical Package for the Social Sciences) program.

## III. FINDINGS AND COMMENTS

Independent variables of the homework method (traditional and online homework practices) and the homework technique (individual and co-operation) attempting to determine the effects on the dependent variable of student success administered immediately after the measurement of success mean scores and standard deviations are given in Table 1.

TABLE I. The Mean and Standard Deviation of Scores

| | | Homework Technique | | |
|---|---|---|---|---|
| | | *Individual* | *Cooperative* | *Total* |
| *Traditional* | *Mean* | 66,00 | 82,07 | 74,31 |
| | *Std. Dev.* | 24,65 | 15,90 | 21,78 |
| | *Participant* | 14 | 15 | 29 |
| *Online* | *Mean* | 88,71 | 67,20 | 77,59 |
| | *Std. Dev.* | 13,34 | 28,69 | 24,78 |
| | *Participant* | 14 | 15 | 29 |
| *Total* | *Mean* | 77,36 | 74,63 | 75,95 |
| | *Std. Dev.* | 22,63 | 24,01 | 23,19 |
| | *Participant* | 28 | 30 | 58 |

Mean and standard deviations of all groups of subjects are presented in Table 1. In order to find out whether there is a statistically significant difference between the means, variance analysis is done.

TABLE II. Comparison of success scores

| | Sum of Sq | Std. Dev. | Mean Sq. | F | p |
|---|---|---|---|---|---|
| **Homework method** | 222,98 | 1 | 222,98 | .476 | **.49** |
| **Homework technique** | 107,45 | 1 | 107,45 | .230 | **.63** |
| **Method * Technique** | 5113,60 | 1 | 5113,60 | 10,93 | **.002\*\*** |
| **Error** | 25276,19 | 54 | 468,08 | | |
| **Total** | **365205,00** | **58** | | | |

According to the variance analysis results on success scores shown in Table II, the value observed for homework preparation method (p=.49) and the value for homework preparation technique variable (p=.63) has no significant difference statistically. In other words, there is no statistically significant difference between the success scores of students participating traditional homework practices, online homework practices and individual homework practices and cooperative homework practices.

However, according to the variance analysis results in Table 2, the value (p=.002) two homework methods (traditional and online; individual and cooperative) consisting of four different homework practices (traditional individual, online individual, traditional cooperative and online cooperative) is statistically significant. Results of Fisher LSD comparison test applied to find out the reason of these difference is given in Table III.

TABLE III. Success scores of Test Results (p values)

|  | Tradition-Individ. | Online–Individ. | Tradition-Cooper. | Online–Cooper. |
|---|---|---|---|---|
| Tradition - Individ. | 1.000 | | | |
| Online-Individ. | .008** | 1.000 | | |
| Tradition - Cooper. | .051 | .412 | 1.000 | |
| Online–Cooper. | .882 | .010** | .065 | 1.000 |

According to Fisher's LSD test, the observed differences in success scores, the traditional cooperative (Tradition_Cooper) success scores of students participating in homework practices, traditional individual (Tradition-Individ.) homework practice (p =. 051), and online individual (Online –Individ.) homework practice (p =. 412) has no statistically significant difference between success scores of the students . Online collaborative (Online – Cooper.) success scores of students participating in homework practices, traditional individual (Tradition-Individ.) homework practice (p =. 882) and the traditional cooperative (Tradition - Cooper.) homework practices (p =. 065) there is no statistically significant difference in the achievement scores.

According to Fisher's LSD test, the observed differences in success scores, online individual (Online-Individ.) homework success scores of students participating in the practice of traditional individual (Tradition-Individ.)  homework success scores of students participating in the implementation statistically (p =. 008) different from the .01 level.  In other words, the success scores of students participating in the online individual practices, is statistically higher than the students participating in the practice of traditional individual practice. The achievement score of students participating Online individual (Online –Individ.) practice is different from scores of students participating online collaborative (Online – Cooper.) statistically (p =. 01) different in .01 level.  In other words, the success scores of students participating in the online practice, is higher than a success score of students participating in online collaborative homework.

IV. RESULTS AND RECOMMENDATIONS

No statistically significant difference is found between the success scores of students participating online homework practices and the traditional one. In the same way, no significant difference between the success scores of students participating individual homework practices and cooperative practices. The academic success of the students participating online-based individual homework, however, is found higher than the success of the students participating online cooperative based homework practices. There is no statistically significant difference between the academic successes of the students participating other 4 practices (traditional individual, online individual, traditional cooperative, online cooperative). Findings of this study developed the following recommendations:

• The teacher candidates can use online individual homework practices to improve their success.

• Homework based on online individual and traditional cooperation practices as developed for this study can be used by the professors.

• The effects of homework based online individual and traditional cooperation practices should be investigated on the success, learning retention, and attitudes of students who will be teachers.

REFERENCES

[1] C.A. Hughes, K.L. Ruhl, J.B. Schumaker and D.D. Deshler. (2002). "Effects of Instruction in an Assignment Completion Strategy on the Homework Performance of Students with Learning Disabilities in General Education Classes," Learning Disabilities Research & Practice, vol 17, Win 2002, pp. 1-18.

[2] D. Muijs and D. Reynolds, Effective Teaching: Evidence and Practice. London: Paul Chapman Publishing, 2001.

[3] D. Kriesberg, "Outdoor Homework," Science Activities, vol. 33, Fall 1996, pp. 23-25.

[4] S. Basaran and B. Tulu, "Asynchronous Education in the Age of Information," Proc. Internet in Turkey Symp. Nov. 1999: Translated in "Bilisim Caginda Asenkron Egitim Aglarının Konumu," V."Türkiye'de İnternet" Konferansı. 19-21 Nov. 1999, Ankara.

[5] M.E. Mutlu, M. C. Ozturk and N. Cetinoz, "Internet Based Training Model, " Open and İDstance Learning Symp. 23-25 May 2002: Translated in "Alternatif Eğitim Araçları İle Zenginleştirilmiş İnternete Dayalı Eğitim Modeli," Acik ve Uzaktan Egitim Sempozyumu 23-25 May 2002.

[6] G.D. Chen, K.L. Ou, C.C. Liu and B.J. Liu, "Intervention and Strategy Analysis for Web Group-Learning," Journal of Computer Assisted Learning, vol 17, 2001, pp. 58-71.

[7] D.W. Johnson and R.T. Johnson, Cooperation and Competition: Theory and Research. Edina, MN: Interaction, 1989.

[8] K. Wilkinson and L. Echternacht, "Internet Homework Activities and Traditional Homework Activities: the Effects on Achievement, Completion Time, and Perception," . Delta Pi Epsilon Journal, vol 40, Fall 1998, pp. 214-30.

[9] M.C. O'Melia and M.S. Rosenberg. "Effects of Cooperative Homework Teams on the Acquisition of Mathematics Skills by Secondary Students with Mild Disabilities," Exceptional Children, vol 60, May 1994, pp. 538-548.

[10] J.L. Epstein and F.L. Van Voorhis, "More Than Minutes: Teachers' Roles in Designing Homework," Educational Psychologist, vol. 36, Summer2001, pp. 181-193.